\documentclass{article}
\usepackage{spconf,amsmath,graphicx}
\usepackage{algorithm}
\usepackage{algorithmic}
\usepackage{pgfplots}

\usepackage{booktabs}
\usepackage{multirow}
\usepackage{colortbl}
\usepackage{diagbox}
\usepackage{float}
\usepackage{makecell}
\usepackage{textcomp}

\usepackage{amsmath}
\usepackage{graphicx}
\usepackage{pgfplots}
\usepackage{pgfplotstable}
\usepackage{caption}

\pgfplotsset{width=7cm,compat=1.3}

\title{INTERRUPTED AND CASCADED PERMUTATION INVARIANT TRAINING FOR SPEECH SEPARATION}

\name{Gene-Ping Yang \qquad Szu-Lin Wu \qquad Yao-Wen Mao \qquad Hung-yi Lee \qquad Lin-shan Lee}
\address{Department of Computer Science and Information Engineering, Nation Taiwan University}
\begin{document}

%
\maketitle
\begin{abstract}

\end{abstract}
Permutation Invariant Training (PIT) has long been a stepping stone method for training speech separation model in handling the label ambiguity problem. With PIT selecting the minimum cost label assignments dynamically, very few studies considered the separation problem to be optimizing both the model parameters and the label assignments, but focused on searching for good model architecture and parameters. In this paper, we investigate instead for a given model architecture the various flexible label assignment strategies for training the model, rather than directly using PIT. Surprisingly, we discover a significant performance boost compared to PIT is possible if the model is trained with fixed label assignments and a good set of labels is chosen. With fixed label training cascaded between two sections of PIT, we achieved the state-of-the-art performance on WSJ0-2mix without changing the model architecture at all.

\vspace{-7pt}

\vspace{-7pt}
\section{Introduction}
\label{sec:intro}
\vspace{-7pt}

Speech Separation has always been a very important issue in speech processing specially in real world application scenarios, in which very often the considered speech signal is disturbed by some additional signals coming from other speakers, so needs to be properly separated. For example, when transcribing meeting verbatim, it was found that people usually speak over the discussions of other people.

Recent advances of deep learning methods have shown outstanding performance on speech separation with good examples including Deep-Clustering based models \cite{hershey2016deep,wang2018alternative,wang2018deep} and Conv-Tasnet \cite{luo2018tasnet}. Most of such approaches first transform the time-domain mixture waveform into some feature map, such as the spectrogram or 2-D feature map encoded by 1-D convolution blocks. An often used approach is then to infer a mask for each individual speaker\cite{wang2014training,luo2018speaker,wang2018end,luo2018tasnet1,isik2016single, chen2017deep}, and multiply the masks element-wise with the mixture feature map to obtain the individual feature maps. A recent work integrating different mixture representations and performing cross-domain joint clustering for mask-inference has also shown encouraging improvements \cite{Yang2019}.

However, these mask-inferring models often suffer from the label ambiguity problem. Suppose there are $T$ mixed utterances in the training set, ${x_i(t), i=1,}$ ${...,T}$, each consisting of $2$ individual signals $x_i(t):[s_{i1}(t), s_{i2}(t)]$. When the machine gives two output signals $y_{i1}(t)$ and $y_{i2}(t)$ for $x_i(t)$, there are two possible label assignments: $[y_{i1}(t) \to s_{i1}(t), y_{i2}(t) \rightarrow s_{i2}(t)]$ and $[y_{i1}(t) \rightarrow s_{i2}(t), y_{i2}(t) \rightarrow s_{i1}(t)]$. For computing the objective function for supervised learning, the label assignments are needed in evaluating the distances between the outputs and the ground truth. This is the label ambiguity problem. There are a total of $2^T$ permutations of the label assignments for all the $T$ mixtures in the training set, or $(N!)^T$ permutations if each mixture includes $N$ signals.

Although Deep Clustering seemed to have avoided this label ambiguity problem by optimizing the similarities between the embeddings of each t-f bin, it turns out that the mask inference branch achieves significantly better performance than the Deep Clustering branch in Chimera++ network \cite{wang2018alternative}. Therefore, for better performance the label ambiguity problem seems not avoidable for the moment. The goal of this paper is to find a good solution to this problem.

Permutation Invariant Training (PIT) has been popularly used to handle the above problem \cite{yu2017permutation,kolbaek2017multitalker}. In this paper we verify experimentally that PIT is not a good solution, because it dynamically assigns the label to each training mixture in an epoch, and such assignments are changed from epoch to epoch. We therefore propose various strategies for more flexible label assignment, and find there can be different ways to do better than PIT.



\vspace{-7pt}
\section{Permutation Invariant Training (PIT) and Its Problems}

Permutation invariant training (PIT) was proposed to handle the label ambiguity problem, in which the loss function for each of the $2$ (or $N!$) label assignments are computed for each mixture signal $x_i(t)$ and the one with the minimum loss is chosen. The model parameters may be updated after seeing every $M$ mixtures based on the $M$ loss functions computed from the minimum loss labels for the $M$ mixtures, and the model updated $T/M$ times in each epoch for the total of $T$ mixtures. In the next epoch the minimum loss label assignment for each mixture will be re-selected again. So PIT adopts dynamically selected rather then fixed label assignment from epoch to epoch.

There exists inevitable problems with PIT. For example, very often in the early stage of training the relatively poor output signals may make the loss values of the $N!$ possible label assignments very close in most of the training mixtures \cite{Yousefi2019}, which means the label assignment may be very random even if they were selected based on the minimum loss criterion. Also, it was found that even after 20 or 30 epochs the minimum loss label assignments for quite high percentage of training mixtures may be reversed in two consecutive epochs, and switched back-and-forth from epoch to epoch, which implies the model parameters may be tuned toward opposite directions repeatedly, or the learning paths may be quite rugged. These observations showed the inadequacy of PIT.

\vspace{-7pt}
\section{Flexible Label Assignment Strategies}
\vspace{-7pt}
Because of the above problems with the dynamic label assignments in PIT, we propose here to make the label assignment more flexible in various ways. A few example strategies are listed below.

\vspace{-7pt}
\subsection{Energy-based Label Assignment}
\vspace{-7pt}

We evaluated the average energy per time frame (with silence automatically detected and deleted) for the two individual signals $s_{i1}(t), s_{i2}(t)$ of each mixture $x_i(t)$. So we can simply assign the higher-energy ground truth to the first model output channel, and the lower-energy ground truth to the other, and this label assignment is fixed throughout all epochs.
\vspace{-7pt}
\subsection{Speaker-embedding-based Label Assignment}
\vspace{-7pt}

We can also extract speaker embeddings (e.g. d-vector\cite{Wan2017GeneralizedEL}, i-vector\cite{5545402} or x-vector\cite{8461375}) for each single speaker utterance $s_{i1}(t)$ or $s_{i2}(t)$ (assuming $N=2$) for all training mixtures $x_i(t)$ in the training set with pre-trained speaker verification models. We then perform a constrained clustering to group all these $T \times 2$ speaker embedding vectors for single speaker utterances in the training set into $2$ clusters, with the constraint that the single speaker utterances which are mixed into a mixture in the training set must have speaker embedding vectors belonging to different clusters, since they are expected to be observed at the $2$ different output channels of the separation model.

Assume $c_1, c_2$ are the two clusters with mean vectors $m_1, m_2$, and $s_1, s_2$ are the embedding vectors of two single speaker utterances $s_{i1}(t), s_{i2}(t)$ of a training mixture. Let $d(s,m)$ denotes the distance between vectors $s$ and $m$. The above constraint can be easily realized by assigning $[s_1 \rightarrow c_1, s_2 \rightarrow c_2]$ if $d(s_1,m_1)+d(s_2,m_2) < d(s_1,m_2)+d(s_2,m_1)$, otherwise the other way, and updating the mean vectors after all single speaker utterances in the training set are assigned. This process can be iterated until converged. With the clustering results, we simply assign the single speaker utterances in cluster $1$ to the first model output channel, and the other to the second, and this assignment is fixed throughout all epochs.

\vspace{-7pt}
\subsection{Fixed Label Assignments Obtained with PIT}
\vspace{-7pt}
In PIT the label assignment for each mixture can be dynamically changed from epoch to epoch, which may be a source of the problem. Here the huge number of assignment permutations are in fact an additional set of unknown parameters to be learned, and as a result with updated model parameters the label assignments can be changed. So we propose to train a model with PIT for $L$ epochs first, and record the label assignments for each mixture at the $L$-th epoch. These label assignments can be considered as good enough labels for training model parameters. So we re-initialize the model parameters and train a new separation model, but with the labels fixed as obtained above with $L$ epochs of PIT.
\vspace{-7pt}
\subsection{Interrupted PIT with Inserted Section of Fixed Label Training}
\vspace{-7pt}
PIT followed by fixed label training as proposed above in Section $3.3$ sounds reasonable, but the set of fixed labels obtained with $L$ epochs of PIT may become inadequate after the model parameters are properly updated by the fixed label training. Therefore, we can perform a new section of PIT again to allow the label assignments to be changed dynamically again after the section of fixed label training. This may solve the poor initialization problem of PIT, that is, during the early stage of training PIT the relatively poor outputs make the label assignments more or less random. This is because here with the fixed label training, the second section of PIT is initialized with a set of much better model parameters, and thus has the potential to further boost the model performance. In this way, the training process actually includes three cascaded sections: (PIT)-(fixed label training)-(PIT), or the PIT process is interrupted after the first section of $L$ epochs and inserted with the second section of fixed label training.


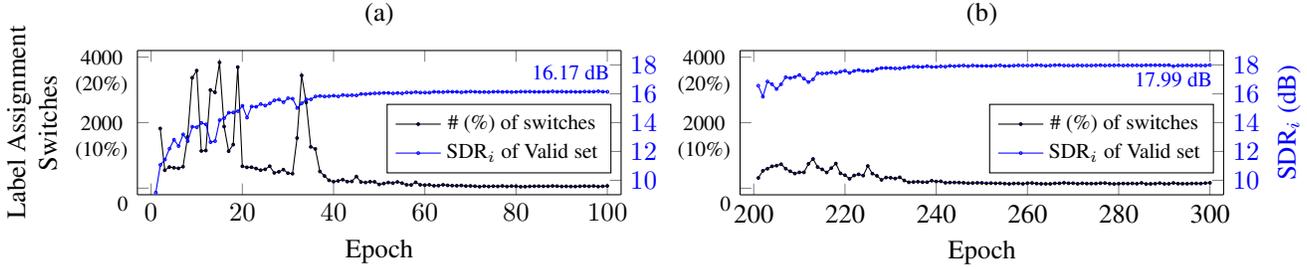
\begin{figure*}
\centering
\begin{tikzpicture}

\begin{axis}[
    title=(a),
    axis y line*=left,
    xlabel={Epoch},
    ymin=-200, ymax = 4200,
    ytick={0,2000,4000},
    yticklabels={0,2000\\(10\%),4000\\(20\%)},
    yticklabel style={align=center, font=\footnotesize, yshift=-0.2cm},
    ylabel style={align=center},
    ylabel={Label Assignment\\ Switches},
    xmax=103, xmin=-3,
    name=first,
    width=.45\linewidth,
    height=3.5cm,
]
\addplot+[color=black, mark=*, mark size=0.5pt,]
    coordinates {(2, 1821)(3, 547)(4, 644)(5, 621)(6, 605)(7, 654)(8, 1564)(9, 3368)(10, 3588)(11, 1135)(12, 1148)(13, 2995)(14, 2924)(15, 3835)(16, 1880)(17, 1128)(18, 1329)(19, 3694)(20, 667)(21, 635)(22, 638)(23, 600)(24, 550)(25, 575)(26, 670)(27, 467)(28, 499)(29, 565)(30, 457)(31, 442)(32, 1529)(33, 3439)(34, 2616)(35, 1260)(36, 1198)(37, 509)(38, 373)(39, 240)(40, 199)(41, 222)(42, 259)(43, 200)(44, 206)(45, 355)(46, 189)(47, 174)(48, 196)(49, 193)(50, 114)(51, 141)(52, 173)(53, 155)(54, 189)(55, 155)(56, 103)(57, 98)(58, 178)(59, 102)(60, 93)(61, 102)(62, 73)(63, 77)(64, 102)(65, 85)(66, 78)(67, 73)(68, 82)(69, 64)(70, 69)(71, 60)(72, 45)(73, 62)(74, 51)(75, 61)(76, 54)(77, 68)(78, 51)(79, 63)(80, 59)(81, 45)(82, 46)(83, 54)(84, 59)(85, 60)(86, 57)(87, 58)(88, 66)(89, 61)(90, 52)(91, 56)(92, 51)(93, 51)(94, 58)(95, 64)(96, 56)(97, 66)(98, 49)(99, 57)(100, 67)
};
\label{p1}
\end{axis}

\begin{axis}[
    axis y line*=right,
    axis x line=none,
    xmax=103, xmin=-3,
    ymax=19, ymin=9,
    width=.45\linewidth,
    height=3.5cm,
    every axis label/.append style = {blue},
    every tick label/.append style = {blue},
    legend style={at={(0.98,0.63)}, font=\footnotesize,}
]
\addlegendimage{/pgfplots/refstyle=p1}\addlegendentry{ \# (\%) of switches}
\addplot+[color=blue, mark=o, mark size=0.5pt,]
    coordinates{(1, 9.147658137702942)(2, 11.070284022235871)(3, 11.432832601547242)(4, 12.195201228237153)(5, 12.81182229309082)(6, 12.370595868873597)(7, 13.18337411327362)(8, 12.7123361287117)(9, 13.714632866287232)(10, 13.677724668502808)(11, 13.978364801025391)(12, 13.879942210388183)(13, 12.642919289779663)(14, 12.715219913864136)(15, 14.171806995773315)(16, 14.306315460586548)(17, 14.687219580078125)(18, 14.70270532836914)(19, 14.79036985168457)(20, 15.165551666641235)(21, 14.348240408515931)(22, 15.117383238601684)(23, 15.09723095741272)(24, 15.285562734222411)(25, 15.184680716323852)(26, 15.334566926193236)(27, 15.545869370269775)(28, 15.601331795501709)(29, 15.412321183013916)(30, 15.69256383304596)(31, 15.655926503753662)(32, 15.014796017837524)(33, 15.337132875823974)(34, 15.608426820373536)(35, 15.612975505828857)(36, 15.8121402885437)(37, 15.84582260055542)(38, 15.81158200340271)(39, 15.830436476898193)(40, 15.867288455963134)(41, 15.819127908706665)(42, 15.913652450180054)(43, 15.87833593902588)(44, 15.9007905254364)(45, 15.886502187919616)(46, 15.988527870178222)(47, 15.974020412826539)(48, 16.00777114868164)(49, 16.030008673858642)(50, 16.035843664550782)(51, 16.058354120635986)(52, 16.05943215713501)(53, 16.02926569786072)(54, 16.055239435958864)(55, 16.028046285247804)(56, 16.091496868896485)(57, 16.076901319503783)(58, 16.11365294570923)(59, 16.08136135673523)(60, 16.08325662536621)(61, 16.08179116668701)(62, 16.13852249221802)(63, 16.118619955825807)(64, 16.113300252914428)(65, 16.14668438606262)(66, 16.13552886543274)(67, 16.136201361083984)(68, 16.101165936660767)(69, 16.131393046951295)(70, 16.154534429168702)(71, 16.133990965652465)(72, 16.11940903816223)(73, 16.122337368011475)(74, 16.11233841018677)(75, 16.13936008720398)(76, 16.1458049911499)(77, 16.12571502532959)(78, 16.138779306793214)(79, 16.117777850341795)(80, 16.12749098892212)(81, 16.14879431819916)(82, 16.14363479385376)(83, 16.149661695861816)(84, 16.13507861442566)(85, 16.130682759475707)(86, 16.139424576568604)(87, 16.134757969284056)(88, 16.16717650756836)(89, 16.139035721969606)(90, 16.153506673812867)(91, 16.13029024925232)(92, 16.14490931739807)(93, 16.131062615966798)(94, 16.13137947998047)(95, 16.15069878349304)(96, 16.14393429374695)(97, 16.143097369384765)(98, 16.175124003219604)(99, 16.143685474014283)(100, 16.13695016784668)}
node at (95,85) {\footnotesize 16.17 dB};
\label{p2}
\addlegendentry{SDR$_i$ of Valid set}
\end{axis}

\begin{axis}[
    title=(b),
    axis y line*=left,
    xlabel={Epoch},
    xtick={0,20,40,60,80,100},
    xticklabels={200, 220, 240,260,280,300},
    xmax=103, xmin=-3,
    ymin=-200, ymax = 4200,
    ytick={0,2000,4000},
    yticklabels={0,2000\\(10\%),4000\\(20\%)},
    yticklabel style={align=center, font=\footnotesize, yshift=-0.2cm},
    width=.45\linewidth,
    height=3.5cm,
    at={(0.45\linewidth,0)}
]
\addplot+[color=black, mark=*, mark size=0.5pt,]
    coordinates {(1, 315)(2, 533)(3, 596)(4, 669)(5, 690)(6, 734)(7, 598)(8, 518)(9, 449)(10, 485)(11, 485)(12, 745)(13, 893)(14, 647)(15, 576)(16, 482)(17, 614)(18, 754)(19, 455)(20, 398)(21, 288)(22, 390)(23, 425)(24, 386)(25, 679)(26, 446)(27, 379)(28, 256)(29, 327)(30, 321)(31, 296)(32, 333)(33, 237)(34, 181)(35, 191)(36, 198)(37, 185)(38, 186)(39, 231)(40, 205)(41, 211)(42, 167)(43, 175)(44, 177)(45, 173)(46, 168)(47, 154)(48, 160)(49, 165)(50, 164)(51, 150)(52, 157)(53, 149)(54, 153)(55, 142)(56, 133)(57, 145)(58, 148)(59, 135)(60, 155)(61, 160)(62, 143)(63, 136)(64, 135)(65, 134)(66, 150)(67, 150)(68, 166)(69, 152)(70, 154)(71, 137)(72, 152)(73, 143)(74, 142)(75, 143)(76, 155)(77, 158)(78, 141)(79, 132)(80, 141)(81, 145)(82, 142)(83, 148)(84, 139)(85, 163)(86, 131)(87, 134)(88, 140)(89, 157)(90, 135)(91, 136)(92, 160)(93, 140)(94, 140)(95, 139)(96, 145)(97, 147)(98, 143)(99, 156)(100, 166)
};
\label{p3}
\end{axis}
\begin{axis}[
    axis y line*=right,
    axis x line=none,
    xmax=103, xmin=-3,
    ymin=9, ymax=19,
    ylabel={SDR$_i$ (dB)},
    y label style={xshift=35pt,yshift=5pt},
    ylabel near ticks,
    every axis label/.append style = {blue},
    every tick label/.append style = {blue},
    legend style={at={(0.98,0.63)}, font=\footnotesize,},
    width=.45\linewidth,
    height=3.5cm,
    at={(0.45\linewidth,0)}
]
\addlegendimage{/pgfplots/refstyle=p3}\addlegendentry{\# (\%) of switches}
\addplot+[color=blue, mark=o, mark size=0.5pt,]
    coordinates{(1, 16.56)(2, 15.8)(3, 16.85)(4, 16.66)(5, 16.33)(6, 16.66)(7, 17.15)(8, 17.09)(9, 17.16)(10, 17.3)(11, 17.05)(12, 16.81)(13, 17.0)(14, 17.42)(15, 17.41)(16, 17.42)(17, 17.49)(18, 17.42)(19, 17.53)(20, 17.6)(21, 17.45)(22, 17.58)(23, 17.65)(24, 17.58)(25, 17.58)(26, 17.59)(27, 17.72)(28, 17.78)(29, 17.79)(30, 17.78)(31, 17.76)(32, 17.76)(33, 17.8)(34, 17.85)(35, 17.89)(36, 17.87)(37, 17.86)(38, 17.91)(39, 17.86)(40, 17.86)(41, 17.9)(42, 17.88)(43, 17.94)(44, 17.91)(45, 17.94)(46, 17.96)(47, 17.91)(48, 17.96)(49, 17.93)(50, 17.93)(51, 17.93)(52, 17.92)(53, 17.95)(54, 17.97)(55, 17.92)(56, 17.93)(57, 17.95)(58, 17.97)(59, 17.96)(60, 17.96)(61, 17.96)(62, 17.99)(63, 17.97)(64, 17.94)(65, 17.96)(66, 17.98)(67, 17.99)(68, 17.96)(69, 17.96)(70, 17.95)(71, 17.97)(72, 17.96)(73, 17.96)(74, 17.96)(75, 17.97)(76, 17.98)(77, 17.96)(78, 17.97)(79, 17.97)(80, 17.98)(81, 17.96)(82, 17.97)(83, 17.97)(84, 17.97)(85, 17.96)(86, 17.98)(87, 17.98)(88, 17.98)(89, 17.98)(90, 17.97)(91, 17.99)(92, 17.91)(93, 17.96)(94, 17.96)(95, 17.97)(96, 17.96)(97, 17.97)(98, 17.96)(99, 17.96)(100, 17.99)}
node at (95,80) {\footnotesize 17.99 dB};
\label{p4}

\addlegendentry{SDR$_i$ of Valid set}
\end{axis}

\end{tikzpicture}
\centering
\vspace{-7pt}

\captionsetup{justification=centering}
\caption{Number (and percentage) of label assignment switches vs. validation SDR$_i$ at each epoch of PIT:\\ (a) PIT with poor model initialization. (b) PIT with good model initialization in the 3$^{rd}$ section of cascade.} 
\label{shift}
\vspace{-7pt}
\end{figure*}

\vspace{-7pt}
\section{Experiments}
\label{sec:pagestyle}
\vspace{-7pt}
\subsection{Experimental Setup} 

We evaluated the proposed approaches on the publicly available dataset WSJ0-2mix \cite{hershey2016deep}, which was derived from WSJ0 corpus. The training objective is to maximize the signal-to-distortion ratio (SDR)\cite{vincent2006performance} of the predicted separated speech $\hat{s}$ against the ground truth $s$,
\begin{equation}
\begin{aligned}
  \operatorname{SDR}( s,\hat{s}) = 10 \log_{10}{\dfrac{\langle s, \hat{s}\rangle^2}{\|s\|^2 \|\hat{s}\|^2-\langle s, \hat{s}\rangle^2}}\,,
  \label{time_loss}
\end{aligned}
\end{equation}
where $\langle \cdot,\cdot \rangle$ represents the dot product and $\|s\|^2=\langle s,s \rangle$ denotes the signal power. The results are primarily reported in SDR improvements, SDR$_i$, which is the SDR values compared to those of the mixture signals against the ground truth.

The goal here is to analyze the training process and label assignments, so we simply utilized the well known second version of Tasnet\cite{luo2018tasnet} previously proposed as the separation model for easier comparison of results.

\vspace{-7pt}
\subsection{Label Assignment Switches for PIT}
\vspace{-7pt}
Here "label assignment switch" refers to the situation that the label assignment of the same mixture was different within two consecutive epochs. We use this to analyze the problems of PIT. We trained a separation model with PIT using the training set. Fig.\ref{shift}(a) shows the total number (and percentage) of the label assignment switches out of the T training mixtures on the left scale at every epoch compared to the immediate previous epoch. It can be seen that there can be thousands (or up to 20\%) of switches and abrupt jumps before epoch 35, for example between epochs 7 and 20, and epochs 31 and 35. Also shown are the SDR$_i$ values achieved with the validation set at each epoch. We can see SDR$_i$ drops very often synchronized with jumps in label switches. It was not until epoch 36 did the SDR$_i$ values rise stably, and the label assignment switch reduce quickly at the same time.

This verified those mentioned earlier that inconsistent label assignments caused unstable training, specially in the early stage of training. This is why the various flexible label assignment strategies mentioned above make sense.

\begin{figure}[h]
    \centering
    \pgfmathsetmacro{\BarOffset}{0.15}
    \pgfplotsset{
    MyAxis/.style={
            scale only axis,
            width=0.66\linewidth,
            xmin=-6,
            xmax=108,
      }
    }
    \pgfplotstableread{
    1  14.55 15.88 1848      
    10 16.76 17.09 692      
    20 16.97 17.26 453      
    30 17.16 17.47 395  
    40 17.23 17.56 206
    50 17.14 17.43 100
    60 17.14 17.47 113
    70 17.18 17.47 62
    80 17.36 17.66 0
    90 16.95 17.27 63
    100 16.93 17.31 47
    }\dataset
    \begin{tikzpicture}
    \begin{axis}[
        ybar,
        bar width=5pt,
        ymax=2000,
        nodes near coords,
        node near coord style={font=\scriptsize},
        ymin=0,
        grid style=dashed,
        MyAxis,
        ylabel style={align=center, yshift=5pt},
        ylabel={\#(\%) of Different Label Assignments \\compared to $L=80$},
        ytick pos=right,
        ytick={0,500,1000,1500,2000},
        yticklabels={0,500\\\footnotesize{(2.5\%)},1000\\(5\%),1500\\\footnotesize{(7.5\%)},2000\\(10\%),},
        yticklabel style={xshift=0.5ex, align=left, font=\small},
        xtick=\empty,
        clip=false,
        tickwidth=5pt,
        xtick style={draw=none}
    ]
    \addplot[color=black, fill=gray!60] table[y index=3] \dataset;
    \end{axis}
    
    \begin{axis}[
        ylabel={SDR$_i$ on validation/testing sets},
        ylabel style={align=center, yshift=-5pt},
        ymax=18,
        ymin=14,
        ymajorgrids=true,
        xtick={1,10,20,...,100},
        xtick pos=left,
        ytick pos=left,
        grid style=dashed,
        yticklabel style={xshift=-0.5ex},
        tickwidth=5pt,
        minor x tick style = {opacity=0},
        MyAxis,
    ]
    \addplot[color=blue, line width=0.5pt, mark=*, mark size=1pt] table[y index=1] \dataset;
    \label{test_sdr}
    \addplot[color=blue, only marks, mark=none, nodes near coords, node near coord style={font=\scriptsize, below=1pt, }] coordinates {
        (80, 17.36) (40, 17.23) (100, 16.93)
        };
    \label{testsdr}
    \addplot [blue,line legend, sharp plot,update limits=false, ] coordinates { (-6,15.82) (108,15.82) }
    node at (65,160) {\footnotesize 15.82 (PIT baseline, test)}
    node at (65,290) {\small test}
    ;
    \end{axis}
    
    \begin{axis}[
        ymax=18,
        ymin=14,
        ymajorgrids=true,
        xtick={1,10,20,...,100},
        xtick pos=left,
        ytick pos=left,
        grid style=dashed,
        yticklabel style={xshift=-0.5ex},
        tickwidth=5pt,
        minor x tick style = {opacity=0},
        MyAxis,
        legend cell align={left},
        legend style={font=\small, at={(0.97,0.35)}},
    ]
    \addplot[color=red, line width=0.5pt, mark=*, mark size=1pt,] table[y index=2] \dataset; 
    \addlegendentry{validation}
    \addlegendimage{/pgfplots/refstyle=test_sdr}\addlegendentry{test}
    \addplot[color=red, only marks, mark=none, nodes near coords, node near coord style={font=\scriptsize},] coordinates {
        (80, 17.66) (40, 17.56) (100, 17.31)
        };
    \addplot [red,line legend, sharp plot,update limits=false, ] coordinates { (-6,16.17) (108,16.17) }
    node at (65,235) {\footnotesize 16.17 (PIT baseline, validation)}
    node at (65,365) {\scriptsize validation}
    ;
    
    \label{validsdr}
    \end{axis}    
    \end{tikzpicture}

    \caption{\small SDR$_i$ achieved on validation/testing sets and number of different label assignments compared to L=80 with fixed label assignments obtained after L epochs of PIT.}
    \vspace{-7pt}
    \label{1to100plot}
\end{figure}
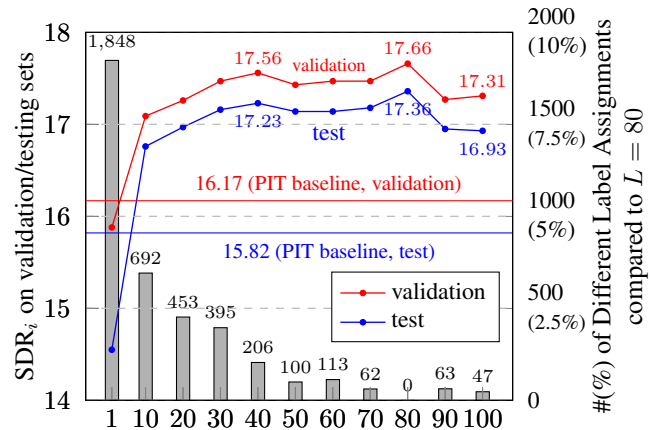

\vspace{-7pt}
\subsection{Fixed Label Training}
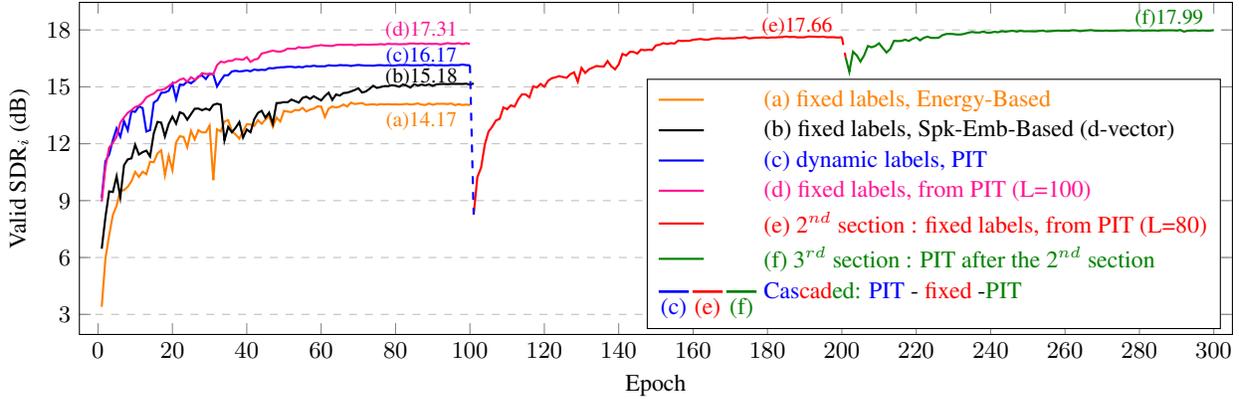
\begin{figure*}[t]
    \centering
    \pgfplotsset{
      /pgfplots/three segment/.style n args={3}{
        legend image code/.code={
            \draw[#1, line width=1pt, no markers] (0cm,0cm) -- node[#1,below,midway] {(c)} (0.4cm,0cm);
            \draw[#2, line width=1pt, no markers] (0.45cm,0cm) -- node[#2,below,midway] {(e)} (0.85cm,0cm);
            \draw[#3, line width=1pt, no markers] (0.9cm,0cm) -- node[#3,below,midway] {(f)} (1.3cm,0cm);
        }
      }
    }
    \pgfplotsset{
      /pgfplots/centered segment/.style={
        legend image code/.code={
            \draw[#1, line width=1pt, no markers] (0.45cm,0cm) -- (0.85cm,0cm);
        }
      }
    }
    \small
    \begin{tikzpicture}
    \centering
    \begin{axis}[
        ytick={3,6,9,12,15,18},
        xlabel=Epoch,
        ylabel=Valid SDR$_i$ (dB),
        legend cell align={left},
        legend style={at={(.99,0.77)}},
        width=0.95\linewidth,
        height=6cm,
        xmin=-5, xmax=305,
        ymajorgrids=true,
        grid style=dashed,
    ]
        \addplot[color={rgb,255:red,255;green,127;blue,0}, mark=none, line width=0.8pt,] table {tas-58-his.tex}
        node at (92,98) {\footnotesize (a)14.17};
        \addlegendentry{\textcolor{rgb,255:red,255;green,127;blue,0,}{(a) fixed labels, Energy-Based}};
        
        \addplot[color=black, mark=none, line width=0.8pt,] table {tas-59-his.tex}
        node at (92,121) {\footnotesize (b)15.18};
        \addlegendentry{\textcolor{black}{(b) fixed labels, Spk-Emb-Based (d-vector)}};
        
        \addplot[color=blue, mark=none, line width=0.8pt,] table {tas-22-his.tex}
        node at (92,132) {\footnotesize (c)16.17};
        \addlegendentry{\textcolor{blue}{(c) dynamic labels, PIT}};
        
        \addplot[color={rgb,255:red,255;green,20;blue,147}, mark=none, line width=0.8pt,] table {tas-51-his.tex}
        node at (92,146) {\footnotesize (d)17.31};
        \addlegendentry{\textcolor{rgb,255:red,255;green,20;blue,147}{(d) fixed labels, from PIT (L=100)}};

        \addplot[color=red, mark=none, line width=0.8pt,] table {tas-49.tex}
        node at (193,148) {\footnotesize (e)17.66};
        \addlegendentry{\textcolor{red}{(e) 2$^{nd}$ section : fixed labels, from PIT (L=80)}};

        \addplot[color=green!50!black, mark=none, line width=0.8pt,] table {tas-49-3.tex}
        node at (293,152) {\footnotesize (f)17.99};
        \addlegendentry{\textcolor{green!50!black}{(f) 3$^{rd}$ section : PIT after the 2$^{nd}$ section}};
        
        \addlegendimage{three segment={blue}{red}{green!50!black}};
        \addlegendentry{
            \textcolor{blue}{Cas}\textcolor{red}{cad}\textcolor{green!50!black}{ed:} \textcolor{blue}{PIT} - \textcolor{red}{fixed} -\textcolor{green!50!black}{PIT}
        };

        \addplot[color=blue, mark=none, dashed, line width=0.8pt,] coordinates {(100, 16.13695016784668) (101, 8.26449831790924)};
        \addplot[color=red, mark=none, dashed, line width=0.8pt,] coordinates {(200, 17.6196931640625) (201, 16.562312712097167)};
    \end{axis}
    
    \end{tikzpicture}
    \vspace{-10pt}
    \centering
    \captionsetup{justification=centering}
    \caption{\small SDR$_{i}$ scores on the validation set at each epoch for different training approaches:\\ (a)(b)(d)(e) fixed labels, (c)(f) dynamic labels (PIT), (e)(f) cascaded 2$^{nd}$ and 3$^{rd}$ sections.}
    \vspace{-7pt}
    \label{cascase_plot}
\end{figure*}

\vspace{-7pt}
Similar SDR$_i$ values obtained with fixed label training with the fixed labels obtained by the energy-based approach mentioned in Section 3.1 are plotted as curve (a) in Fig.\ref{cascase_plot}, similarly by the speaker-embedding-based approach mentioned in Section 3.2 as curve (b) in Fig.\ref{cascase_plot}. Both of them are significantly lower (converged to 14.17 and 15.18 dB respectively) than that obtained with PIT in curve (c) of Fig.\ref{cascase_plot}, which is exactly the same curve in Fig.1(a) converged to 16.17 dB. This shows fixed labels alone were inadequate, and PIT is clearly better even with unstable training due to serious label switches.

So we next tested a different way of obtaining the fixed labels, by PIT after $L$ epochs as mentioned in Section 3.3. The results for $L=1,10,20,30,...,100$ are depicted in Fig.\ref{1to100plot}, where the red and blue curves are respectively for validation and test sets, all with 100 epochs of training. We can see as long as $L\geq 10$ the training converged to SDR$_i$ values significantly higher than the baseline of 16.17 dB achieved by PIT (curve (c) in Fig.\ref{cascase_plot}), with best result of 17.66 dB achieved at $L=80$, for the validation set. The two curves for validation and test are in general parallel.

The vertical bars in Fig.\ref{1to100plot} are the number of label assignments out of the $T$ training mixtures which were different from that for $L=80$. We see for $L=90$ or $100$ only a small number of labels were different, but these different labels made differences in the finally converged SDR$_i$ values.

\vspace{-7pt}

\subsection{PIT cascaded with Fixed Label Training }
\vspace{-7pt}
Curves (d)(e) of Fig.\ref{cascase_plot} are for fixed label training, with the labels fixed at those obtained with $L=100$ and $L=80$ epochs of PIT respectively. Here because the first 100 epochs were performed with PIT to obtain the fixed labels, and the next 100 epochs (101 to 200) were performed with fixed label training. This is why this curve (e) is plotted over 101 to 200 epochs and converged to 17.66 dB, which is the value at $L=80$ in Fig.\ref{1to100plot}. Curve (d) (actually epoch 101 to 200) is plotted over 1 to 100 epochs only for better comparison with those from other methods. 


As mentioned in Section 3.4, we can performed an additional section of PIT at the end of curve (e), allowing the labels to be changed for another 100 epochs, or over epochs 201 to 300. The result is curve (f) of Fig.\ref{cascase_plot}, converged at 17.99 dB for SDR$_i$. The cascaded three sections of (PIT)-(fixed label training)-(PIT) is actually curves (c)(e)(f) in Fig.\ref{cascase_plot}, giving a result 1.82 dB higher than PIT ($17.99-16.17=1.82$).

The number (and percentage) of label assignment switches for the third section of PIT, or curve (f) for epochs 201 to 300 in Fig.\ref{cascase_plot}, are also plotted in Fig.1(b), to be compared with those in Fig.1(a) for the first section of PIT, or curve (c) over epochs 1-100 in Fig.\ref{cascase_plot}, together with the SDR$_i$ values in Fig.\ref{cascase_plot}. Both Fig.1(a)(b) are for 100 epochs of PIT, except Fig.1(a) started with a very poor model so with large numbers and abrupt jumps of label assignment switches, while Fig.1(b) started with a well-trained model, so with well reduced and smoothed label assignment switches.

\subsection{Summary of the Results}

We summarize the different approaches analyzed here in Table 1, in which the rows labeled by (a)-(f) correspond to curves (a)-(f) in Fig.3. The column of "Labels" indicated whether the label assignments are fixed, dynamic (dyn), or with cascaded (csc) sections of PIT and fixed labels. The SDR$_i$ values are those finally converged to for validation and test sets, and the last column are the percentages of labels out of all the T training mixtures for which the fixed labels or finally obtained labels are different from the best results (epoch 300 at the end of curve (f)). We see by cascading with a section of fixed labels the validation SDR$_i$ was improved from 16.17 dB of PIT (100 epoch) in row (c) to 17.99 dB of cascade of three sections (300 epoch) in row (f). We also see the high correlation between the SDR$_i$ values and percentages of different labels in the last two columns verifying the point here.

We also compare the results with those of prior works in Table 2, all data on the test set for comparison, including scale-invariant signal-to-noise ratio improvement (SI-SNR$_i$) \cite{luo2018speaker} and SDR$_i$ in dB. Row (a) is for the baseline separation model TasNet-v2 \cite{luo2018tasnet} we used throughout this work. Rows (b)(c) are respectively the latest version of TasNet and our implementation using Prob-PIT\cite{Yousefi2019} with TasNet-v2. Row (d) is for our previous work of cross-domain joint clustering which was not used here at all. Row (e)(f) are for the approaches proposed here, corresponding to curves (e)(f) in Fig.\ref{cascase_plot}. We see decent improvement by the proposed approach (cascaded 3 sections) even without using the cross-domain approach in row(d).

\vspace{-7pt}
\begin{table}
    \caption{\small Summary of all approaches analyzed here. Row (a)-(f) corresponds to curves (a)-(f) in Fig.3. SDR$_i$ are the finally converged values on the validation set. The last column are the percentages of label assignment which are different from that in row (f).}    
    \vspace{-7pt}
    \small
    \centering
    \begin{tabular}{|c|c|c|c|c|c|}
        \Xhline{2\arrayrulewidth}
        \rowcolor{gray!20}
        \multicolumn{2}{|c|}{}&&&&\\
        \rowcolor{gray!20}
        \multicolumn{2}{|c|}{\multirow{-2}{*}{\textbf{Approaches}}} & 
        \multirow{-2}{*}{\textbf{Labels}} & 
        \multirow{-2}{*}{\shortstack[c]{\textbf{Valid} \\\textbf{SDR$_i$}}} & 
        \multirow{-2}{*}{\shortstack[c]{\textbf{Test}\\\textbf{SDR$_i$}}} & 
        \multirow{-2}{*}{{\shortstack[c]{\textbf{\% Diff}\\\textbf{Labels}}}} \\
        \Xhline{2\arrayrulewidth}
        \multicolumn{2}{|l|}{(a) Energy-based} & \multirow{4}{*}{\shortstack[c]{fixed}} & 14.17 & 14.09 & 48.8\% \\
        \cline{1-2} \cline{4-6}
        \multirow{3}{*}{\shortstack[c]{Spk-Emb\\based}} & (b) d-vector & & 15.18 & 14.88 & 43.1\% \\
        \cline{2-2} \cline{4-6}
        & x-vector & & 14.35 & 13.87 & 41.5\% \\
        \cline{2-2} \cline{4-6}
        & i-vector & & 14.50 & 14.10 & 41.4\% \\
        \hline
        \multicolumn{2}{|l|}{(c) PIT} & dyn & 16.17 & 15.82 & 1.4\% \\
        \hline
        \multicolumn{2}{|l|}{(d) fx from PIT (L=100)} & fixed & 17.31 & 16.93 & 1.4\% \\
        \hline
        \multicolumn{2}{|l|}{(e) fx from PIT (L=80)}  & \multirow{2}{*}{\shortstack[c]{csc}} & 17.66 & 17.36 & 1.4\% \\
        \cline{1-2} \cline{4-6}
        \multicolumn{2}{|l|}{(f) (PIT)-(fx)-(PIT)} &  & 17.99 & 17.74 & 0\% \\
        \Xhline{2\arrayrulewidth}
    \end{tabular}
    \vspace{-7pt}
    \label{proposed1}
\end{table}

\begin{table}[t]
  \caption{\small $\text{SI-SNR}_i$ and $\text{SDR}_i$ (in test set) compared to different prior works tested on WSJ0-2mix dataset. "*" indicates our implementation not written in the original paper.}
  \label{prior}
  \centering
  \small
  \begin{tabular}{|c|l|c|c|c|}
    \Xhline{2\arrayrulewidth}
    \rowcolor{gray!20}
    \multicolumn{2}{|c|}{\textbf{Approaches}} & \textbf{Params} &  \textbf{$\text{SI-SNR}_i$} &  \textbf{$\text{SDR}_i$} \\
    \Xhline{2\arrayrulewidth}
    \multirow{4}{*}{\rotatebox[origin=c]{90}{prior works}}
    & (a) TasNet-v2 \cite{luo2018tasnet} &  8.8{\scriptsize M} & 14.6 {\scriptsize dB} & 15.0 {\scriptsize dB} \\
    & (b) Conv-TasNet \cite{luo2019conv} &  5.1{\scriptsize M} & 15.3 {\scriptsize dB} & 15.6 {\scriptsize dB} \\
    & (c) Prob-PIT \cite{Yousefi2019} &  8.8{\scriptsize M} & 15.9$^*$  {\scriptsize dB} & 16.2$^*$ {\scriptsize dB} \\
    & (d) Yang et al. \cite{Yang2019} &  10{\scriptsize M}  & 16.6 {\scriptsize dB}  & 16.9 {\scriptsize dB} \\
    \hline
    \multicolumn{2}{|l|}{(e) \small{Cascaded:} (PIT)-(fx)} &  8.8{\scriptsize M}  & 17.1 {\scriptsize dB}  & 17.4 {\scriptsize dB} \\
    \multicolumn{2}{|l|}{(f) \small{Cascaded:} (PIT)-(fx)-(PIT)} &  8.8{\scriptsize M}  & 17.5 {\scriptsize dB}  & 17.7 {\scriptsize dB} \\
    \Xhline{2\arrayrulewidth}
  \end{tabular}
\end{table}

\vspace{-7pt}
\section{Conclusion}
\vspace{-7pt}
In this paper, we propose to train a separation model by interrupted and cascaded PIT with a fixed label training section inserted in the middle, whose label assignments are obtained by the first section of PIT training. State-of-the-art performance of SDR$_i = 17.7$ dB was achieved on the WSJ0-2mix test set. This verified that the label assignments obtained by PIT are good for fixed label training, and a well trained model is also beneficial for further PIT training.

\bibliographystyle{IEEEbib}
\bibliography{strings,refs}

\end{document}